\documentclass[twocolumn,prl,aps,floats,superscriptaddress]{revtex4}
\usepackage{graphicx}
\usepackage{epsfig}
\usepackage{epsf}

\def\Tel{{\cal T}_{el}}
\def\chiel{\chi_{el}}
\def\Tcel{T_c^{el}}
\def\Dcel{D_c^{el}}
\def\vR{{\bf R}}
\def\Dtrel{D_{tr}^{el}(T)}
\def\vtrel{v_{tr}^{el}(T)}

\def\v2o3{V$_2$O$_3$\,}
\def\vcr2o3{(V$_{1-x}$Cr$_x$)$_2$O$_3$\,}
\def\vo2{VO$_2$\,}

\def\srvo3{SrVO$_3$\,}
\def\cavo3{CaVO$_3$\,}
\def\latio3{LaTiO$_3$\,}
\def\lasrtio3{La$_{1-x}$Sr$_x$TiO$_3$\,}
\def\ytio3{YTiO$_3$\,}

\def\ceal3{CeAl$_3$\,}
\def\naxcoo2{Na$_x$CoO$_2$\,}
\def\kappacl{$\kappa$-(BEDT-TTF)$_{2}$Cu[N(CN)$_{2}$]Cl\,}

\def\kappaX{$\kappa$-(BEDT-TTF)$_{2}$X\,}

\begin{document}

\title{Sound Velocity Anomaly at the Mott Transition: \\
application to organic conductors and V$_2$O$_3$}

\author{S.R.Hassan}
\affiliation{Laboratoire de Physique Th\'eorique-Ecole Normale Superieure\\
24,rue Lhomond 75231 Paris Cedex 05, France}
\affiliation{Centre de Physique Th\'eorique-Ecole Polytechnique, 91128 Palaiseau Cedex, France}
\author{A.Georges}
\affiliation{Centre de Physique Th\'eorique-Ecole Polytechnique, 91128 Palaiseau Cedex, France}
\affiliation{Laboratoire de Physique Th\'eorique-Ecole Normale Superieure\\
24,rue Lhomond 75231 Paris Cedex 05, France}
\author{H.R.Krishnamurthy}
\affiliation{Centre for Condensed Matter Theory, Department of
Physics, Indian Institute of Science, Bangalore 560 012, India}
\affiliation{Condensed Matter Theory Unit, JNCASR, Jakkur,
Bangalore 560 064, India.}

\begin{abstract}
Close to the Mott transition, lattice degrees of freedom react
to the softening of electron degrees of freedom. This results in a
change of lattice spacing, a diverging compressibility and a
critical anomaly of the sound velocity. These effects are
investigated within a simple model, in the framework of dynamical
mean-field theory. The results
compare favorably to recent experiments on the layered organic
conductor \kappacl. We predict that effects of a similar magnitude
are expected for V$_2$O$_3$, despite the much larger value of the
elastic modulus of this material.
\end{abstract}

\maketitle

The Mott transition, which is the metal-insulator transition (MIT)
induced by electron-electron interactions, has been investigated
theoretically and experimentally for many
years~\cite{Mott,imada_mit_review}. Some materials are poised very
close to the Mott transition, which can therefore be induced by
varying pressure, temperature, or chemical composition. This is
the case of \vcr2o3, and of the family of layered molecular
crystals \kappaX where X is an anion
(e.g.,X=I$_3$,Cu[N(CN)$_{2}$]Cl,Cu(SCN)$_2$). In these compounds,
one observes a pressure induced, finite temperature, first order
phase transition. Pressure increases the bandwidth, reducing the
relative interaction strength. The first-order transition line
ends at a second-order critical endpoint $(P_c,T_c)$. The critical
behaviour at this endpoint has been recently the subject of
remarkable experimental
investigations~\cite{limelette_v2o3_science,kagawa_bedt}.

On the theory side, our understanding of the Mott transition has
benefitted from the development of dynamical mean-field theory
(DMFT)~\cite{AGeorges,Mott_brief}. In this theory, electronic degrees of
freedom are the driving force of the transition, and the critical
endpoint is associated with a diverging electronic response
function $\chiel$ (defined below)~\cite{HRK,rozenberg_finiteT_mott}.
An analogy exists with the
liquid-gas transition. The insulating phase is a low-density
gas of neutral bound pairs of doubly occupied and empty sites; the
metal is a high-density liquid of unbound doubly- occupied and
empty sites which therefore conduct.
The scalar order parameter is associated with the low-energy
spectral weight.

In real materials however, lattice degrees of freedom do play a
role at the Mott transition. This is expected on a physical basis:
in the metallic phase the itinerant electrons participate more in
the cohesion of the solid than in the insulating phase where they
are localized. As a result, the lattice spacing increases when
going from the metal to the insulator. A discontinuous change of
the lattice spacing through the first-order metal-insulator
transition line is indeed observed in \vcr2o3~\cite{jayaraman}.
Raman scattering experiments~\cite{Ylin} find that in the metallic
state the frequency of certain phonons associated with the
BEDT-TTF molecules has a non-monotonic temperature dependence
below 200 K. The effect of the Mott transition on optical
(Einstein) phonons was studied theoretically in~\cite{Mckenzie}.
Acoustic experiments~\cite{Poirier1} find an anomaly in the sound
velocity of the organic materials as a function of temperature,
with a particularly dramatic reduction recently
observed~\cite{Poirier2} for \kappacl near the Mott critical
endpoint at $T_c\simeq 40K$.

In this paper, we propose a simple theory of the effects connected
with lattice expansion through the Mott transition, and the
associated divergence of the compressibility. We address in
particular the critical anomaly of the sound velocity observed in
acoustic experiments. For this purpose, both the electronic
degrees of freedom and the ionic positions must be retained in a
model description. We adopt the simplest possible framework,
previously introduced in Ref.\cite{HRK}, namely the compressible
Hubbard model (see also~\cite{cyrot_1972_compress}), with the
electron correlations being treated within DMFT. Our results
compare favorably to the recent acoustic experiments on the
layered organic conductor \kappacl [referred to in the rest of
this paper simply as the organic-conductor(OC).] Furthermore, we
show that effects of a similar magnitude are expected for \v2o3,
despite the much larger value of the elastic modulus of this
material.

In the following, we assume that the dependence of the free energy
on the unit-cell volume $v$ (or, rather on the induced strain, see below)
can be written as:
\begin{equation}\label{eq:free}
F= F_0 - P_0 (v-v_0) +
\frac{1}{2}\,B_0{\frac{(v-v_0)^2}{v_0}+F_{el}[D(v)]}
\end{equation}
The last term $F_{el}$, is the contribution of the electronic
degrees of freedom which are active through the transition (e.g
the d-shell for \v2o3). Specifically we take $F_{el}$ to be the
free energy of a single band, half-filled Hubbard
model~\cite{footnote} with a half-bandwidth $D(v)$ depending on
the unit-cell volume $v$. The first three terms arise from
expanding the free energy due to other degrees of freedom about a
reference cell volume $v_0$, $P_0$ and $B_0$ being the
corresponding (reference) pressure and the bulk elastic modulus.
Expression (\ref{eq:free}) can be derived from a microscopic
hamiltonian
$H=H_{lat}[\vR_i]-\sum_{ij\sigma}t(\vR_i-\vR_j)d^\dagger_{i\sigma}d_{j\sigma}
+U\sum_i n_{i\uparrow}n_{i\downarrow}$, involving both the ion
positions $\vR_i$ and the electronic degrees of freedom, when all
phonon {\it excitations} are neglected, i.e all lattice
displacements $\vR_i-\vR_j$ are taken to be uniform. It is
conventional (see e.g~\cite{Harrison}) to use an exponential
parametrization for $D(v)$, which we linearize since relative
changes in $v$ are small:
$D(v)=D_0\exp{[-\gamma(v-v_0)/v_0]}\simeq
D_0\left[1-\gamma(v-v_0)/v_0\right]$. As a result, the pressure $P
= - {\partial F}/{\partial v}$ and the ``compressibility'' $K \equiv
-(v\partial P/\partial v)^{-1} =(v\partial^2 F/\partial v^2)^{-1}$
are given by:
\begin{eqnarray}
&P=P_0-{B_0}(v-v_0)/{v_0} - ({\gamma D_0}/{v_0})\Tel
\label{eq:pressure}\\
&(K\,v)^{-1}={B_0}/{v_0} - ({\gamma D_0}/{v_0})^2\,\chiel
\label{eq:compress}
\end{eqnarray}
Here $\Tel$ is the (dimensionless) electronic kinetic energy
$\Tel \left(T, D(v),U\right) \equiv -{\partial F_{el}}/{\partial
D}$, and $\chiel$ is the electronic response function:
$\chiel \left(T, D(v),U\right) \equiv
-{\partial^2F_{el}}/{\partial D^2}$.
Both quantities are associated with the purely electronic Hubbard model.
\begin{figure}
\epsfig {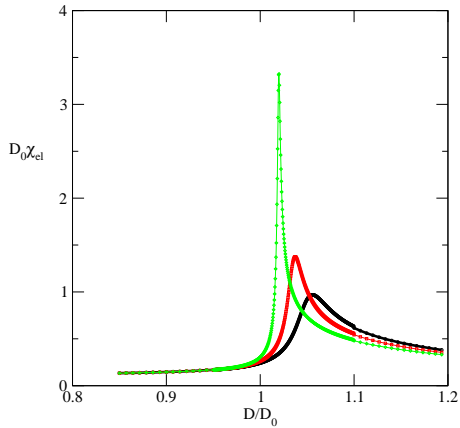} \caption{The variation of
the electronic response function $\chiel$ with the bandwidth $D$
for various temperatures.  The parameters are $U/D_0 = $ 2.492 and
$ T/D_0 = $ 0.060, 0.055, 0.050  from right to left. The results
are from DMFT solved in the IPT approximation.}
\end{figure}
Within DMFT, $\chiel$ is found~\cite{HRK,rozenberg_finiteT_mott}
to reach a peak value $\chiel^{max}(T,U)$ for a given temperature
$T$ at a specific value of  $D = D_m^{el}(T,U)$ , and eventually,
to diverge at $T=\Tcel,\,D=\Dcel$. This is illustrated in Fig.1
using DMFT together with iterated perturbation theory
(IPT)~\cite{AGeorges} in which case
$\Tcel \simeq 0.02\,U,\,\Dcel\simeq 0.4\,U$.
In the compressible model, as pointed out in
Ref.\cite{HRK}, the Mott critical endpoint will therefore occur at
$T_c  > \Tcel $ and will be signalled by the divergence of $K$.
From (\ref{eq:compress}), this happens when $\chiel^{max}$ has
reached a large enough (but not divergent) value such that
$D_0\,\chiel^{max}(T_c) = {B_0v_0}/({\gamma^2 D_0})$. The
corresponding critical half-bandwidth is $D_c =  D_m^{el}(T_c,U)$
and the critical cell volume is determined from $ (v_c-v_0)/{v_0}
= -(D_c-D_0)/(\gamma {D_0})$. The critical pressure can then be
determined using Eq.~(\ref{eq:pressure}). If  $T_c$ is close enough
to $T_c^{el}$ one can use the form (for $T>\Tcel$):
$\chiel^{max} \simeq \alpha_{el}(U)\Tcel/(T-\Tcel)$, which follows from
mean-field theory~\cite{rozenberg_finiteT_mott}. Here,
$ \alpha_{el}(U) = \alpha / U $ with $\alpha\simeq 0.5$
in our calculations. Hence one obtains:
$\Delta T_c/T_c\equiv
({T_c-\Tcel})/{\Tcel}~=~\alpha~[\gamma^2{D_0}/({B_0v_0})](D_0/U).$
%
\begin{figure}
\epsfig{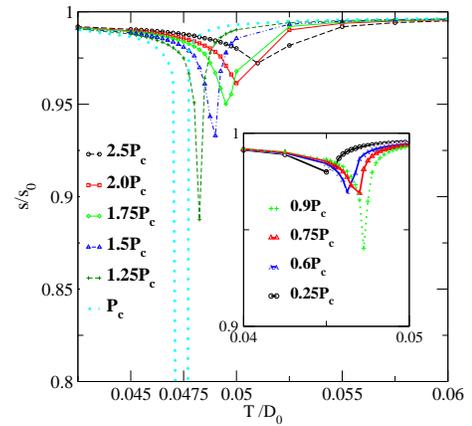} \caption{Temperature
dependence of the sound velocity at various pressures for
parameters corresponding to OC. Inset: position and amplitude of
anomaly below $P_c$}
\end{figure}

The inverse ``compressibility'' $K ^{-1}$ is directly proportional to
the square of the sound velocity $s$: $s \propto 1/ \sqrt{K}$. At
$T_c$, since $K$ {\it diverges}, the
sound-velocity {\it vanishes}. Hence, right at the critical point
the acoustic phonon branch under consideration disperses
anomalously. If the crystal has inversion symmetry, the dispersion
should go as $\omega \propto q^2$. Thus from a calculation of the
inverse compressibility along lines similar to that in Ref.~\cite{HRK}
one can determine the dependence of the sound velocity
on temperature and pressure, including its critical anomaly, as
reported below. Note that a similar critical behaviour of the
sound-velocity is well known in the context of the usual
liquid-gas transition~\cite{sound_liquid_gas}.

We note that the ``pressure'' and inverse ``compressibility'' in
Eqs.~(\ref{eq:pressure},\ref{eq:compress}) are associated with the stiffness
with respect to changes in the unit-cell volume, i.e. to
bulk strain at {\it fixed number} of ions and electrons per unit cell.
General thermodynamics relates $\rho^2\partial\mu/\partial\rho$
(with $\rho$ the total density) to the inverse compressibility provided the
latter is defined from volume changes {\it at zero strain} (i.e stemming from
vacancy diffusion).
Hence, this relation does not apply to $K$ (which is related to the
stress-stress correlation function rather than to the density-density one).
This also implies that $K$ is not related to the ``charge compressibility''
$\kappa_{el}=\partial n_{el}/\partial\mu_{el}$ (studied
theoretically, eg. in \cite{kotliar_compress}),
as the latter involves changes of the
electron density under conditions of zero strain. The
correspondence between the sound velocity and the ``compressibility''
K as calculated above can be shown to be exact for the simplest
compressible Hubbard model of phonons modulating the nearest
neighbour hopping on a cubic lattice, and for longitudinal sound
propagation along the [111] direction. We hence
present it as a reasonable zeroth- order description of real
systems. A more realistic description of anisotropic materials
should take into account the dependence on both polarization and direction
of propagation.
\begin{figure}
\epsfig{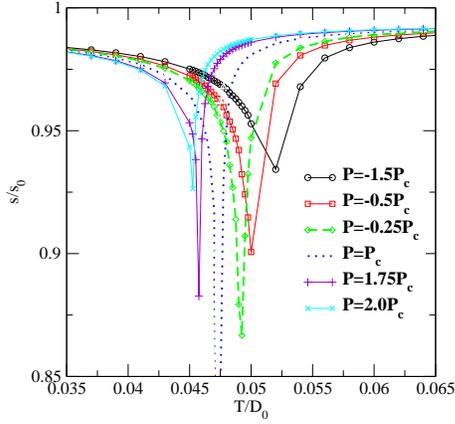} \caption{Temperature
dependence of the sound velocity at various pressures for
parameters corresponding to pure V$_2$O$_3$.
``Negative'' pressures can be reached by
chromium-substitutions:
for \vcr2o3, experiments establish that $x=1\%$ corresponds to
$\Delta P=-4$kbar ($\simeq P_c$) (see e.g~\cite{imada_mit_review}).}
\end{figure}
\begin{table}[tbp] \centering
\begin{tabular}
[c]{|l|l|l|}
\hline \textbf{Parameter} & $V_{2}O_{3}$ & $OC$\\
\hline $D_{0}$ & 1 eV &.13 eV\\
\hline $v_0$ & 100 $\mathring{A}^{3}$ & 1700 $\mathring{A}^{3}$\\
\hline $B_{0}$ & 2140 kbar & 122 kbar\\
\hline $\gamma$ & 3 & 5\\
\hline $B_{0}v_{0}$ & 133 eV & 129 eV\\
\hline $U/D_0$ & 2.468  & 2.492 \\
\hline $B_{0}v_{0}/(\gamma^2D_0)$ & $14.7$ & $40$\\
\hline
\end{tabular}
\caption{Table of parameter values for \v2o3 and the
organic-conductor \kappacl (OC)} \label{table1}
\end{table}

The details of the calculational procedure have been described in
Ref.~\cite{HRK}, and we avoid repeating them here. The parameter
values we have chosen for the two materials are given in Table~1.
Our approach is to consider pure \v2o3 and the OC at ambient
pressure (i.e., $P_0 =$ 1 bar is essentially zero) as reference
compounds, for which $v_0$ and $B_0$ are taken from
experiments.
For the OC, we take the values $D_0\simeq 0.13$~eV and $\gamma=5$,
which were found
in~\cite{limelette_bedt_prl} to be consistent with transport data.
For \v2o3, we take $D_0=1$~eV and $\gamma=3$ (values
considered standard for d-electron systems~\cite{Harrison}).
Finally, the value of $U/D_0$ is adjusted so that the critical
pressure is reproduced correctly ($P_c\simeq -4$~kbar for \v2o3,
$P_c\simeq +200$~bar for the OC). This requires $U/D_0$ to be
poised very close to the Mott critical value for the pure
electronic problem (1.26 in our calculations). Using the
parameters in Table.~1, one sees that the (dimension-less)
combination $B_0v_0/(\gamma^2D_0)$ is large for both compounds
($\simeq 14.7$ for \v2o3 and $\simeq 40$ for the OC).
This implies that the Mott transition arises for large values of
$D_0\chiel$, i.e., the experimentally observed transition {\it is
definitely driven by the electronic degrees of freedom}, and very
close to the purely electronic Mott transition. Specifically,
we find
that the relative shift $\Delta\,T_c/T_c$ of the critical temperature due to the coupling
to the lattice, is as small as $1.4 \%$ for \v2o3 and even smaller for
the organic conductors. In fact it is remarkable that despite the
very different values of the bulk modulus of the two materials
(the OC being much softer than \v2o3), the combination
$B_0v_0/(\gamma^2D_0)$ only differs by a factor of 3 between them.
As a result, the order of magnitude of the sound-velocity anomaly
is expected to be similar in both materials, since $K_0v_0/K v =
1-(\gamma^2D_0/B_0v_0)\,D_0\chiel$. We emphasize that there is a
significant difference between our choice of parameters and that
made in Ref.~\cite{HRK} for \v2o3: there, a much smaller value of
$B_0$ (much smaller than the measured experimental value) and
somewhat larger values of $U/D$ were used in order to obtain
volume jumps comparable to what is seen experimentally, resulting
in a large relative shift $\Delta T_c/T_c$, on the scale of
$40\%$. Because the contribution of the electronic degrees of
freedom to the total bulk modulus is comparatively small, it is
hard to reconcile the choice made in Ref.~\cite{HRK}
with experimental data for this
quantity. On the other hand, with our choice of $B_0$,
the calculated fractional volume jump $\Delta v /v_0$ through the
transition is $\sim 0.2 \%$, too small compared to the
observed~\cite{jayaraman} jump ($\sim 1 \%$). We comment on a
possible resolution of this problem towards the end of this paper.
\begin{figure}
\epsfig{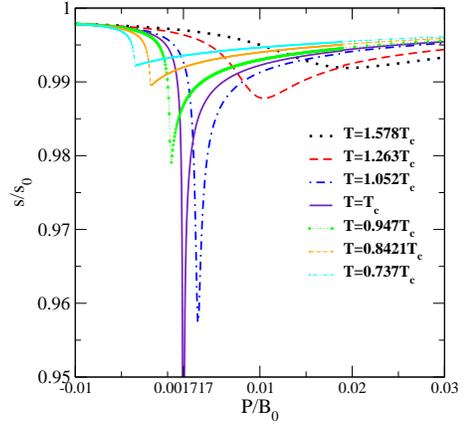} \caption{The sound velocity
as a function of pressure at various fixed temperature for
parameters corresponding to OC}
\end{figure}

Figs. 2 and 3 show our results for the temperature- dependence of
the sound velocity computed using the method described above, for
the parameters representative of the OC and \v2o3 and for several
pressures. At the critical pressure $P=P_c$, the sound-velocity
vanishes according to the mean-field law $s\propto(T-T_c)^{1/2}$
($K$ diverges as $1/(T-T_c)$, as follows from Eq.~(\ref{eq:compress})).
A pronounced dip
remains visible in a rather extended range of pressure both above
and below $P_c$. The overall shape of these curves, as well as the
typical order of magnitude of the effect are in quite good
agreement with the experimental data for the OC, recently
published in~\cite{Poirier2}. For example, we obtain a dip of
relative size $\Delta s/s_0\simeq 10\%$ for $P\simeq 1.3 P_c$,
consistent with the experimental observation.

In Fig.4 and 5, we show the sound- velocity as a function of
pressure for various temperatures. This has been studied in a less
systematic manner in the experiments on the OC, but the overall
shape and magnitude of our results are again consistent with the
published data~\cite{Poirier2}. In particular, the curves in
Figs.4-5 show a marked asymmetry: on the low-pressure (insulating)
side the pressure dependence is rather weak and a dip appears only
very close to $P_c$, while a more gradual and sizeable pressure-
dependence is observed on the high- pressure (metallic) side. This
reflects the asymmetry in the electronic response function
$\chiel$ observed on Fig.1. It is expected qualitatively, since
electrons participate much less to the cohesive energy on the
insulating side, and is also observed
experimentally~\cite{Poirier2}.

The inset of Fig.~5  shows the temperature-pressure phase diagram
for parameters corresponding to \v2o3. The slope $dP_{tr}/dT$
obtained is too large compared to the observed value. We believe
that this, as well as the problem mentioned above, namely the
smallness of the calculated $\Delta v /v_0$, can be resolved as
follows.
\begin{figure}
\epsfig{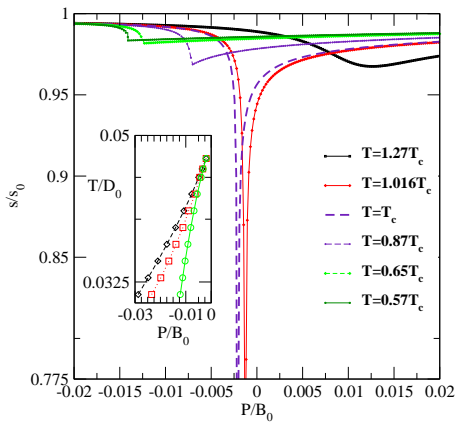} \caption{The sound
velocity as a function of pressure at various fixed temperatures
for parameters corresponding to $V_2O_3$. Inset: The corresponding
phase diagram.
The central line and the lines on either side correspond to the
transition pressure $P_{tr}$ and the spinodal pressures
respectively}
\end{figure}
Let $\Dtrel = D(\vtrel)$ be the half-bandwidth for metal-insulator
coexistence in the purely electronic problem at temperature $T$.
For $T$ well below $ T_c$,
the fractional differential volumes $\delta v_{i,m} /v_0 \equiv
(v_{i,m} - \vtrel)/v_0 $  of the coexisting insulating and
metallic phases and the transition pressure $P_{tr}$ in the
presence of lattice coupling are approximately $\delta v_i / v_0 \simeq
- \delta v_m / v_0 \simeq \gamma ({\cal T}_m - {\cal T}_i)/(
2B_0v_0 / D_0)$ and $P_{tr} \simeq B_0(\Dtrel-D_0)/(\gamma {D_0})-
(\gamma D_0 /v_0)({\cal T}_m + {\cal T}_i)/ 2 $ (compare
\ref{eq:pressure}). Here
${\cal T}_{i,m}$ are $\Tel$  evaluated in the coexisting
insulating and metallic phases respectively. These expressions are
in good agreement with the numerical results reported above and in
Ref.~\cite{HRK}. In the context of more realistic models for the
electronic problem, one expects that differential screening
effects would reduce (enhance) the effective Hubbard $U$ and
correspondingly enhance (reduce) ${\cal T}$, in the metallic
(insulating) phase, leading to an overall enhancement of $(v_i
-v_m)/v_0$. An additional enhancement factor might come from the
orbital degeneracy in \v2o3. These effects could be addressed in
future studies of multi-band
Hubbard models, including screening effects in a self-consistent
manner. High- precision experimental measurements of optical
spectral weights in the coexisting insulating and metallic phases
would also be very interesting, and could provide information on
the kinetic energies ${\cal T}_{i,m}$.


\acknowledgments{We acknowledge discussions with S.~Florens, L.~de Medici,
D.~Jerome, E.~Kats, G.~Kotliar,
P.~Limelette, R.~McKenzie, M.~Poirier, A.M.~Tremblay, P.~Wzietek and
with M.~Rozenberg (who also provided some of his codes).
This work has been
supported by an Indo-French collaborative grant (IFCPAR-2404-1), by
CNRS and by Ecole Polytechnique.}


\begin{thebibliography}{99}
\bibitem{footnote} This is appropriate for the organics. For \v2o3, a more
quantitative description (beyond the scope of this paper) would
require the consideration of orbital degeneracy and two
d-electrons.
\end{thebibliography}
\end{document}